\documentstyle[12pt]{article}

\begin{document}
\renewcommand{\thefootnote}{\fnsymbol{footnote}}
\begin{center}
{\large COMMENT ON CHERNAVSKAYA'S PAPER ``DOUBLE PHASE TRANSITION MODEL AND THE
PROBLEM OF ENTROPY AND BARYON NUMBER CONSERVATION" hep-ph/9701265}\vspace{.4in}\\
{\large V. J. Menon $^1$ and B. K. Patra$^2$\footnote[2]{E-mail: bkpatra@iopb.stpbh.soft.net} } \vspace{.2in}\\ 
{\it {$^1$}Department of Physics, Banaras Hindu University, Varanasi 221 005,
India\\
$^2$Institute of Physics, Sachivalaya Marg, Bhubaneswar 751 005, India}
\end{center}

\section{\bf Introduction}

In the standard construction of a first-order equilibrium phase transition via
the Gibbs criteria from a quark-gluon plasma (QGP)
to a hot and dense hadron gas (HG), the appearance of a discontinuity in the
entropy per baryon ratio (s/n) makes the phase transition at fixed
temperature T and fixed chemical potential $\mu$ irreversible. 
Recently several papers have been addressed to the
question of conserving the entropy per baryon s/n across the phase boundary. 
Leonidov et al. [1] have proposed a bag model equation of state (EOS) for the QGP
consisting of massless, free gas of quarks and gluons using a ($\mu$, T)
dependent bag constant $B(\mu, T)$ in an isentropic equilibrium phase transition
from a QGP to the HG at a constant T and $\mu$. Later Patra and Singh [2] have 
extended this idea to remedy some anomalous behaviour of such a bag constant 
$B(\mu,T)$ through the inclusion of 
perturbative QCD corrections in the EOS for QGP. They have also explored the consequences
of such a bag constant 
on the deconfining phase transition in the relativistic heavy-ion collisions as well as in the early Universe case [3].\\
	The above mentioned analysis refers only to the stationary systems. 
But in the context of modern experiments on ultrarelativistic heavy-ion collisions, 
the dynamical evolution of the system within the framework of hydrodynamical 
models has to be incorporated also. Recently Chernavskaya [4] have suggested
a double phase transition model via 
an intermediate phase containing massive constituent quarks and pions.
He claimed that only the continuity 
condition in s/n ratio does not provide equilibrium character of first-order
transition in dynamically evolving systems 
as for all equilibrium processes, but enthalpy
of the system as well. 
Subsequently he has also criticized the work of Ref. 1 and 2 on
the ground of twin constraints arising from Gibbs - Duham equilibrium relation [5] 
and enthalpy conservation [6] for an evolving system. The aim of the present
note is to logically counter both these criticisms below.\vspace{0.22 in}\\
\section{\bf Gibbs - Duham Relation}
Ref.[4] suggests that if $\epsilon$ is the energy density and p the pressure
then form-invariance of the relation
\begin{eqnarray}
\epsilon + p - \mu n = s T
\end{eqnarray}
imposes the condition
\begin{eqnarray}
\mu ~~ \frac{\partial B}{\partial \mu} = - T ~~ \frac{\partial B}{\partial T}
\end{eqnarray}
We comment that eq.(2) is untenable due to two reason. Firstly, it
would imply that B, instead of being a function of two independent variables
$\mu $ and T, will depend only on the single variable $\mu $/T as can be 
verified by direct differentiation. Secondly, eq.(2) would make it impossible
to apply the iterative analytical procedure [1,2] of solving the basic
partial differential equation based on s/n in the extreme regions 
$\mu \rightarrow $ 0, T $\rightarrow \infty $ as well as $\mu \rightarrow
\infty $, T $\rightarrow$ 0. This would imply a conflict with the QCD
sum rule results [7]. \vspace{.2in}\\
\section{\bf Enthalpy Condition}
Ref.[4] mentions that, if $\omega$ = $\epsilon + P $ is the enthalpy 
density, then for an evolving hydrodynamic system containing the mixed phase of the
QGP and hadronic gas, the conservation of enthalpy per baryon $\omega$/n
gives an additional constraint. We comment that this constraint is 
redundant, i.e., does not give a new information. This is so because even if the
system is evolving with time, we can sit in the local comoving frame where the
relation (1) is expected to hold. Then
\begin{eqnarray}
\frac{\omega}{n} = \frac{(\epsilon + P)}{n} = T ~~ \frac{s}{n} + \mu
\end{eqnarray}
implying that, for any given T and $\mu$, the conservations of $\omega$/n
and s/n are equivalent.\\

\pagebreak

\end{document}